\documentclass[prd,nofootinbib,showpacs,twocolumn]{revtex4-1}
\usepackage{color}
\usepackage{bm}
\usepackage{graphicx}
\usepackage{amsmath}
\usepackage{amssymb}
\usepackage{enumitem}
\usepackage{hyperref}
\usepackage{array}
\usepackage[english]{babel}
\usepackage{tensor}
\usepackage{comment}
\usepackage[normalem]{ulem}
\usepackage[dvipsnames]{xcolor}
\allowdisplaybreaks

\def\be{\begin{equation}}
\def\ee{\end{equation}}
\def\ba{\begin{eqnarray}}
\def\ea{\end{eqnarray}}
\def\l{\left}
\def\r{\right}
\def\f{\frac}

\usepackage{tikz}

\def\nn{\nonumber}

\def\hub{{\mathcal H}}

\definecolor{orange}{rgb}{1,0.5,0}
\definecolor{darkorange}{rgb}{0.69,0.33,0.13}
\definecolor{darkgreen}{rgb}{0.05,0.5,0.06}

\newcommand{\Pobs}{P_{\mathrm{obs}}}

\newcommand{\Tstrut}{\rule{0pt}{2.6ex}}       

\newcommand{\beq}{\begin{equation}}
\newcommand{\eeq}{\end{equation}}
\newcommand{\beqa}{\begin{eqnarray}}
\newcommand{\eeqa}{\end{eqnarray}}

\begin{document}

\title{The road ahead of Horndeski: cosmology of surviving scalar-tensor theories}

\author{Noemi Frusciante$^{1}$, Simone Peirone$^{2}$, Santiago Casas$^{3}$,  Nelson A. Lima$^{4}$}
\affiliation{
\smallskip
$^{1}$ Instituto de Astrof\'isica e Ci\^encias do Espa\c{c}o, Faculdade de Ci\^encias da Universidade de Lisboa,  
 Edificio C8, Campo Grande, P-1749016, Lisboa, Portugal \\
\smallskip
$^{2}$ Institute Lorentz, Leiden University, PO Box 9506, Leiden 2300 RA, The Netherlands\\
\smallskip 
$^{3}$ AIM, CEA, CNRS, Universit\'{e} Paris-Saclay, Universit\'{e} Paris Diderot; Sorbonne Paris Cit\`{e}, F-91191 Gif-sur-Yvette, France \\
\smallskip 
$^{4}$ ITP, Ruprecht-Karls-Universität Heidelberg, Philosophenweg 16, 69120 Heidelberg, Germany 
\smallskip 
}

\begin{abstract}

In the context of the effective field theory of dark energy (EFT) we perform agnostic explorations of Horndeski gravity.
We choose two parametrizations for the free EFT functions, namely a power law  and a dark energy density-like behaviour on a non trivial Chevallier-Polarski-Linder background.  We restrict our analysis to those EFT functions which do not modify the speed of propagation of gravitational waves. Among those, we prove that one specific function cannot be constrained by data, since its contribution to the observables is below the cosmic variance, although we show it has a relevant role in defining the viable parameter space. We place constraints on the parameters of these models combining measurements from present day cosmological datasets and we prove that the next generation galaxy surveys can improve such constraints by one  order of magnitude. We then verify the validity of the quasi-static limit within the sound horizon of the dark field, by looking at the  phenomenological functions $\mu$ and $\Sigma$, associated respectively to clustering and lensing potentials. Furthermore, we notice up to $5\%$ deviations in $\mu, \Sigma$  with respect to General Relativity at scales smaller than the Compton one.
For the chosen parametrizations and in the quasi-static limit, future constraints on $\mu$ and $\Sigma$ can reach the $1\%$ level and will allow us to discriminate between certain models at more than $3\sigma$, provided the present best-fit values remain.
\end{abstract}


\date{\today}

\maketitle

\section{Introduction}\label{Intro}

The attempt to find a definite theory of gravity able  to explain the late time acceleration of the Universe has resulted in a wide selection of dark energy (DE) and modified gravity (MG) models~\cite{Silvestri:2009hh,Clifton:2011jh,Copeland:2006wr,Lue:2004rj,Joyce:2014kja,Tsujikawa:2010zza}.
When exploring the cosmology of these models, it is very useful to employ an unified approach to describe in a model independent fashion any departure from General Relativity (GR). 
Among the many approaches presented in the literature, a popular framework is the one based on the $\mu, \Sigma$ parametrization~\cite{Bertschinger:2008zb,Pogosian:2010tj}, according to which deviations from GR in the Poisson and lensing equations are encoded respectively in the $\mu$ and $\Sigma$ phenomenological functions. However, one has to rely on the quasi-static (QS) approximation in order to express these functions in an analytical form for a chosen theory. For this reason the approach has a limitation given by the break down scale of the QS assumption. Such scale, usually identified with the cosmological horizon, has been claimed to be instead the sound horizon of the dark field~\cite{Sawicki:2015zya}. 

Another general framework, encompassing theories with one additional scalar degree of freedom (DoF), is the Effective Field Theory of dark energy (EFT)~\cite{Gubitosi:2012hu,Bloomfield:2012ff}. Such description parametrises the evolution of linear cosmological perturbations in terms of few free functions of time, dubbed EFT functions. 
The benefit of using the EFT approach relies in a direct connection with the underlying theory of gravity. Indeed, each EFT function multiplies a specific geometrical operator in the action: thus picking out a set of EFT functions translates in selecting a class of DE/MG models. Moreover, the mapping procedure, which allows to translate a specific theory in the EFT language, does not rely on any QS approximation~\cite{Gubitosi:2012hu,Bloomfield:2012ff,Bloomfield:2013efa,Gleyzes:2013ooa,Gleyzes:2014rba,Frusciante:2015maa,Frusciante:2016xoj}.  A resemble of the EFT functions is the $\alpha$-basis~\cite{Bellini:2014fua,Gleyzes:2014qga,Frusciante:2016xoj}. In the latter the free functions  can be directly related to some phenomenological aspects of the DE field, such as the running of the Planck mass, braiding and kineticity effects and deviation in the speed of propagation of tensor modes~\cite{Bellini:2014fua}.  

In the present work, we perform a cosmological investigation by means of agnostic parametrizations in terms of the EFT functions. We  select the subset of EFT functions describing the Horndeski theory~\cite{Horndeski:1974wa} (or Generalized Galileon~\cite{Deffayet:2009mn}).  In particular, we  consider the class of models satisfying the condition $c_t^2=1$, which accommodates the stringent bound on the speed of propagation of tensor modes placed by \textit{LIGO}/\textit{VIRGO} collaborations after  the detection of the gravitational wave (GW) event GW170817 and its optical counterpart~\cite{TheLIGOScientific:2017qsa,Monitor:2017mdv,Coulter:2017wya}. The implication of this result on modified gravity theories has been discussed in several works~\cite{Creminelli:2017sry,Ezquiaga:2017ekz,Baker:2017hug,Sakstein:2017xjx,Bettoni:2016mij,Kase:2018aps} and, in the case of Horndeski, the survived viable action involves a reduced number of free functions~\cite{Creminelli:2017sry}. In particular, the Quintic Lagrangian is removed and the coupling with the Ricci scalar in the Quartic Lagrangian reduces to a general function of the scalar field.  Hereafter, we refer to such action as the ``surviving'' Horndeski action (sH).  Very recently it has been shown that it is possible to build a class of theories where the GWs speed is set to unity dynamically when the scalar is decoupled from the matter sector~\cite{Copeland:2018yuh}.
However, it is worth to notice that the applicability of the GWs constraint is still subject of debate since, as pointed out in ref.~\cite{deRham:2018red}, the energy scales detected by \textit{LIGO} lie very close to the typical cutoff of many DE models.

In the next decade several large scale surveys, such as \textit{DESI}, \textit{Euclid}, \textit{SKA} and \textit{LSST}, are planned to start and they will cover the entire redshift range over which dark energy played a significant role in the accelerated expansion.  Looking forward to having real data, forecasts analysis is improving our knowledge of cosmology by looking both at specific gravity models as well as model-independent parametrizations~\cite{Leung:2016xli,Alonso:2016suf,Casas:2017eob,Heneka:2018ins,Mancini:2018qtb}. 
In this work we provide cosmological constraints on sH theories using both present datasets and future spectroscopic Galaxy Clustering (GC) and Weak Lensing (WL) observables. We show how the latter are able to set tighter constraints on the parameters entering the sH action.

The manuscript is organised as follows. In Sec.~\ref{Sec:theory}, we make an overview of the sH theory and its parametrizations in the EFT formalism. In Sec.~\ref{Sec:Method}, we introduce the agnostic parametrizations defining the sH models, the codes and datasets used for the Monte Carlo Markov Chain analysis as well as WL and GC forecasts. In Sec.~\ref{Sec:Results}, we discuss the results and present the constraints on the models parameters from present and future surveys. Finally, we conclude in Sec.~\ref{Sec:conclusion}.        

\section{Theory}\label{Sec:theory}

\subsection{Horndeski theory and its parametrizations}\label{Horndeskiparameterization}

Horndeski theory has become very popular, as it is the most general scalar tensor theory in four dimensions constructed from the metric $g_{\mu\nu}$, the scalar field $\phi$ and their derivatives, giving second order field equations. Its generality relies in a certain number of free functions in the action, namely $\{\mathcal{K}, G_3,G_4,G_5\}[\phi,X]$, where $X=\partial^\mu\phi\partial_\mu\phi$. The number of these functions was reduced after the detection of the GW170817 event. 
 Indeed, the stringent constraint on the speed of propagation of the tensor modes, disfavours  the presence of the $G_5$ term and reduces $G_4$ to be solely a function of the scalar field~\cite{Creminelli:2017sry}.   Thus, the sH action, which assumes an unmodified speed of propagation of gravitational waves $(c_t^2=1)$, takes the following form:
\be\label{sHaction}
\mathcal{S}_{sH}=\int{}d^4x\sqrt{-g}\l[\mathcal{K}(\phi,X)+G_3(\phi,X)\Box\phi+G_4(\phi)R\r]\,,
\ee  
where $g$ is the determinant of the metric $g_{\mu\nu}$ and $R$ is the Ricci scalar. Despite the Horndeski action drastically simplifies, a high degree of freedom in choosing the above functions still remains.

We are interested in investigating the linear cosmological perturbations, thus in the following we focus on a complementary framework to describe the sH action, i.e.~the EFT approach~\cite{Gubitosi:2012hu,Bloomfield:2012ff}. Within this framework we can write the corresponding  linear perturbed action around a flat Friedmann-Lema$\hat{\text{i}}$tre-Robertson-Walker (FLRW) background and in unitary gauge, which reads 
\begin{align}\label{eftaction}
\mathcal{S} =& \int d^4x \sqrt{-g}  \bigg\{ \frac{m_0^2}{2} \left[1+\Omega(a)\right] R + \Lambda(a) - c(a)\,a^2\delta g^{00}  \nonumber\\
  & + m_0^2H_0^2 \frac{\gamma_1 (a)}{2} \left( a^2\delta g^{00} \right)^2
 - m_0^2H_0 \frac{\gamma_2(a)}{2} \, a^2\delta g^{00}\,\delta K\bigg\}, 
\end{align}
where $m_0^2$ is the Planck mass, $\delta g^{00}$ and $\delta K$,  are  the perturbations respectively of the upper time-time component of the metric and the trace  of the extrinsic curvature, $H_0$ is the Hubble parameter at present time and $a$ is the scale factor. $\{\Omega,c, \Lambda,\gamma_1,\gamma_2\}$ are the so called EFT functions. $\Lambda$ and $c$ can be expressed in terms of $\Omega$, the conformal Hubble function, $\hub$ and the densities and pressures of matter fluids by using the background field equations~\cite{Gubitosi:2012hu,Bloomfield:2012ff}. Thus, we are left with only three free EFT functions. While $\Omega$ acts at both background and perturbations level, $\gamma_1$ and $\gamma_2$ contribute only to the linear perturbations evolution. 

The EFT functions can be specified for a chosen theory once the mapping has been worked out~\cite{Gubitosi:2012hu,Bloomfield:2012ff,Bloomfield:2013efa,Gleyzes:2013ooa,Gleyzes:2014rba,Frusciante:2015maa,Frusciante:2016xoj}. For action~(\ref{sHaction}) the mapping simply reads 
\ba\label{EFT_functions}
1+\Omega&=&\f{2}{m_0^2}G_4\,,\nn\\
m_0^2H_0^2 \gamma_1&=&\mathcal{K}_{XX}X^2-3\f{\hub}{a^6} G_{3XX}\dot{\phi}^5-G_{3\phi X}\f{\dot{\phi}^4}{2a^4}\nn\\
&+&G_{3X}\f{\dot{\phi}^2}{2a^4}\l(\ddot{\phi}+2\hub\dot{\phi}\r)\,,\nn\\
m_0^2H_0\gamma_2&=&-2G_{3X}\f{\dot{\phi}^3}{a^3}\,,
\ea
where dots are derivatives with respect to conformal time, $\tau$ and the subscripts $X$ and $\phi$ are respectively the derivatives with respect to $X$ and $\phi$. Therefore, the EFT approach practically translates the problem of choosing appropriate forms for the $\mathcal{K},G_i$-functions into choosing specific forms of the EFT functions. 

Let us now comment on the functional dependence of the $\mathcal{K}, G_i$-functions. All of them can modify the expansion history regardless of their specific dependence on $\phi$ or $X$. However this is not true at the level of perturbations. In the following there are some examples:
\begin{itemize}
\item  \textit{$G_3$-function}. \textit{$G_3=G_3(\phi)$}: it solely affects the expansion history in the form of a dynamical DE. Indeed, it can be recast as an equivalent contribution of $\mathcal{K}$ in the form $\mathcal{K}=F(\phi)X$ by integration by parts (being $F\propto G_{3\phi}$)~\cite{Deffayet:2010qz}.   
\noindent
         \textit{$G_3=G_3(\phi,X)$}: this function gives a non vanishing $\gamma_1$ and $\gamma_2$. Note that if $\gamma_2\neq 0$ the function $\gamma_1$ is forced to be not zero from eq.~(\ref{EFT_functions})  (except in the case of a fine tuning). The opposite does not hold.  This is an important aspect when selecting the combinations of non-vanishing EFT functions. Finally, $G_{3X} \neq 0$ has been identified to be responsible for the  braiding effect or mixing of the kinetic terms of the scalar and metric~\cite{Deffayet:2010qz}. For this reason, $\gamma_2$ can be interpreted as a braiding function. Thus, in order to parametrise for e.g. the so called Kinetic Gravity Braiding models (KGB)~\cite{Deffayet:2010qz} both $\gamma_1$ and $\gamma_2$ need to be active.   

\item \textit{$G_4$-function}. When $G_4 \neq m_0^2/2$, it is the only function which can  modify the coupling, i.e. $\Omega\neq 0$.  The function $m_0^2(1+\Omega)$ can be interpreted as  an  effective Planck mass and its evolution rate can be defined as $\alpha_M=\dot{\Omega}/\hub(1+\Omega)$~\cite{Bellini:2014fua}.  A running Planck mass  contributes also to the braiding effect: in particular, in the case  $G_{3X} = 0$, the running Planck mass is the sole responsible for the braiding effect~\cite{Bellini:2014fua}. 

\item  \textit{$\mathcal{K}$-function}.   When $\mathcal{K}$  is only a function of $\phi$, it does not give any contribution to the perturbations: in fact $\gamma_1$ does not depend on $\mathcal{K}(\phi)$.  On the contrary when \textit{$\mathcal{K} = \mathcal{K}(\phi,X)$}, it contributes both to the background equations and to the perturbations through $\gamma_1$ (the latter if $\mathcal{K}_{XX}\neq 0$). In particular, in the case $\{G_4(\phi), G_3=0, \mathcal{K}(X)\}$ and $\mathcal{K}_{XX}\neq 0$, the form of $\gamma_1$ is fixed in terms of background functions as $\gamma_1=\f{c}{m_0^2H_0^2}\l(\f{\dot{c}}{\dot{\Lambda}}-1\r)$.

\noindent
In the regime in which the QS approximation holds, it has been found that $\gamma_1$ is negligible for linear cosmological perturbations~\cite{Bloomfield:2012ff,Bloomfield:2013efa}. In Sec.~\ref{Sec:Model} we will show that although $\gamma_1$ is unlikely to be constrained by cosmological data, it still plays a relevant role in defining the stable parameter space of the theory.
\end{itemize}

In order to study the cosmological signatures of each EFT function we introduce the $\mu, \Sigma$ parametrization, which allows to encode all possible deviations from GR at the level of the linear perturbed field equations~\cite{Bertschinger:2008zb,Pogosian:2010tj}. They are defined, respectively, as the deviations from the GR Poisson equation and the GR lensing equation and, in Fourier space, they read 
\ba \label{mudef}
&&-k^2\psi=4\pi G_N a^2\mu(a,k)\rho\Delta\,, \nn\\
&&-k^2(\psi+\phi)=8\pi G_Na^2\Sigma(a,k)\rho\Delta\,,
\ea
where $\{\psi(t,x^i), \phi(t,x^i)\}$ are the gravitational potentials, $G_N$ is the Newtonian gravitational constant and $\rho\Delta=\sum_i \rho_i\Delta_i$, includes the contributions of all fluid components. GR is recovered for $\mu=\Sigma=1$.

Although their definition is very general, explicit and analytical expressions for them can be found considering a specific Lagrangian describing a chosen gravity theory with  one extra scalar DoF, in the QS approximation~\cite{Bellini:2014fua,Silvestri:2013ne}. In such approximation and for the case under analysis they read:
\ba
\mu(a,k)&=& \f{1}{1+\Omega}\, \,\f{1+M^2\f{a^2}{k^2}}{g_1+M^2\f{a^2}{k^2}}\,,\nn \\
\Sigma(a,k)&=& \f{1}{2(1+\Omega)}\, \,\f{1+g_2+M^2\f{a^2}{k^2}}{g_1+M^2\f{a^2}{k^2}}\,, \label{QS_expressions}
\ea
where $g_i$ and $M$ are functions of $a$ and can be expressed in terms of EFT functions, i.e. $\Omega$ and $\gamma_2$.  As anticipated before, $\gamma_1$ does not enter in these expressions because they have been derived in the QS approximation (see ref.~\cite{Silvestri:2013ne} for their explicit expressions and a general discussion, here we address the specific case $c_t^2=1$). $M$ represents the mass of the dark field and, from eq.~(\ref{QS_expressions}), we see that it is responsible for the scale dependence of the phenomenological functions: it defines a new scale associated to the extra DoF, i.e. the Compton scale ($\lambda_C \sim 1/M$).  In the super-Compton limit, i.e. $k/a \ll M$ (subscript ``0''), one gets $\mu_0=1/(1+\Omega)$, $\Sigma_0=\mu_0$. In this limit, the only signature of modification to gravity comes from the coupling function $\Omega$. Such function impact the clustering and  lensing potentials and have effects on the  Cosmic Microwave Background (CMB) lensing and  galaxy weak lensing. Additionally, because of the  late time Integrated Sachs-Wolfe (ISW) effect, it affects  the amplitude of the  low-multipole CMB anisotropies. Finally, because of stability conditions (i.e.~avoidance of ghost instability for tensor modes~\cite{Frusciante:2016xoj}), we have $1+\Omega>0$ thus both $\mu_0$ and $\Sigma_0$ are positive. In the sub-Compton limit (subscript ``$\infty$'') both the expressions involve $\gamma_2$ and $\Omega$.  As in the previous case, $\mu_\infty$ and $\Sigma_\infty$ are modified and if $\gamma_2\neq 0$ it follows $\mu_{\infty} \neq \Sigma_{\infty}$. In this case the effects on the observables are the same as in the previous limit but they are the results of the combination of both $\Omega$ and $\gamma_2$.  At these scales, the gravitational slip parameter, $\eta=2\Sigma/\mu-1$, is modified only if $\gamma_2\neq 0$, allowing for the presence of an anisotropic stress term  related to the viscosity of a DE fluid~\cite{Pujolas:2011he}. On the other hand, if $\gamma_2=0$ it follows $\mu_{\infty}=\Sigma_{\infty}$ and $\eta_\infty=1$.  For stability requirements~\cite{Bellini:2014fua}  $\mu_\infty$ is positive, while a conclusion about $\Sigma_\infty$ is not straightforward. 
In this regard, it has been shown in~\cite{Peirone:2017ywi} that $(\mu-1)(\Sigma-1)\ge0$. 

In Sec.~\ref{Sec:Results} we verify the applicability of the QS approximation within the sound horizon of the dark field for the specific models analysed in this work.

\section{Method}\label{Sec:Method}

\subsection{Models}\label{Sec:Model}

\begin{figure}[ht!]
\begin{center}
\includegraphics[width=0.5\textwidth]{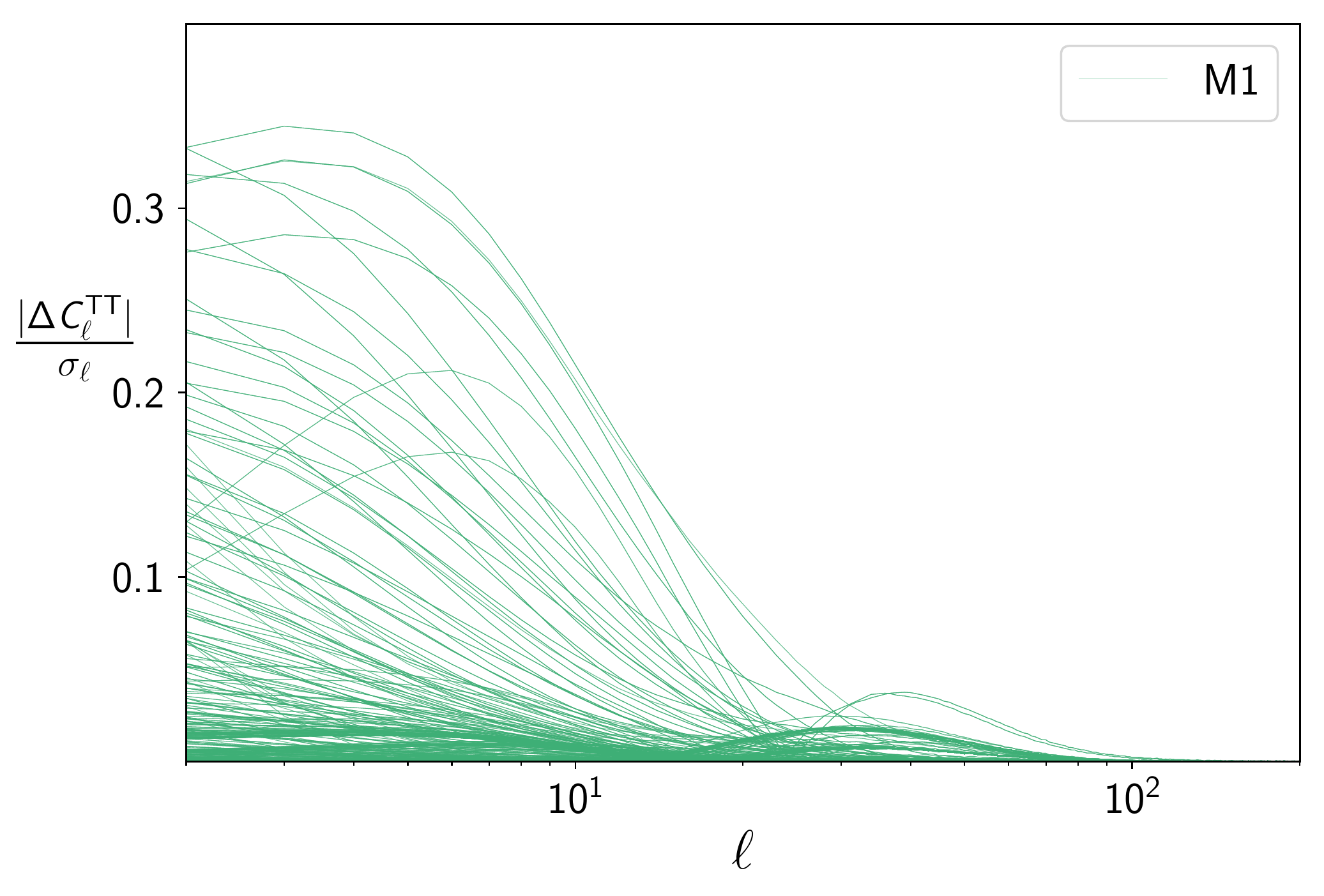}
\caption{\label{Cl_M1} Effects of $\gamma_1$ in {\bf M1} on the TT power spectrum. We plot the deviation on the $C^{\rm TT}(\ell)$, in units of cosmic variance $\sigma_\ell = \sqrt{2/(2 \ell +1)} C^{\rm TT}(\ell)$. We consider a sample of $\sim 10^3$ models, where both $\Omega$ and $\gamma_1$ are parametrised as in {\bf M1}. $\Delta C^{\rm TT}$ is obtained as the difference between the model with $\Omega+\gamma_1$ and the one with solely $\Omega$.}
\end{center}
\end{figure}
\begin{figure}[t!]
\includegraphics[width=0.5\textwidth]{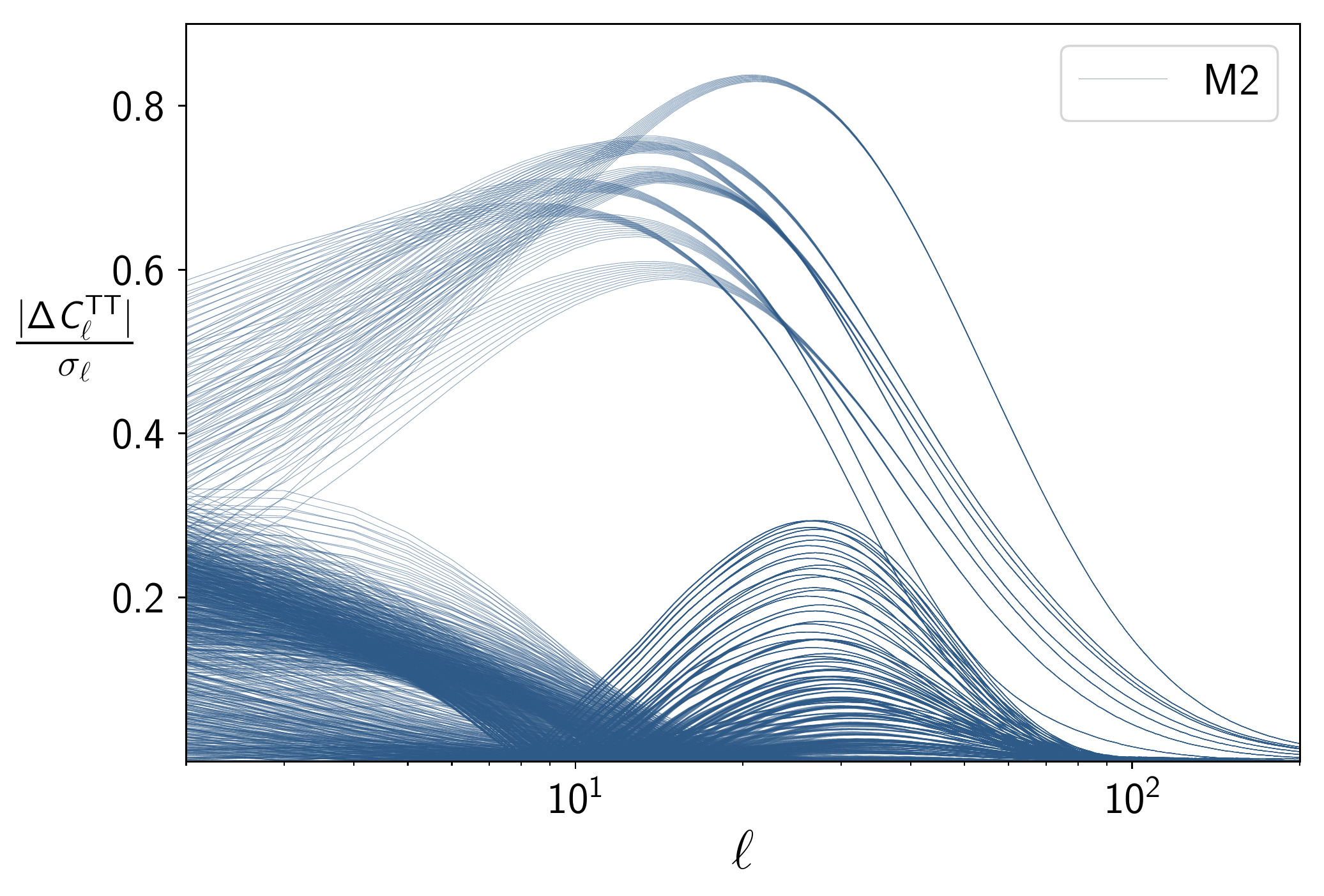}
\caption{\label{Cl_M2} Effects of $\gamma_1$ in {\bf M2} on the TT power spectrum. We plot the deviation on the $C^{\rm TT}(\ell)$, in units of cosmic variance $\sigma_\ell = \sqrt{2/(2 \ell +1)} C^{\rm TT}(\ell)$.  We consider a sample of $\sim 10^3$ models, where both $\Omega$ and $\gamma_1$ are parametrised as in {\bf M2}. $\Delta C^{\rm TT}$ is obtained as the difference between the model with $\Omega+\gamma_1$ and the one with solely $\Omega$.
}
\end{figure}

In this Section we will  present two agnostic parametrizations of the EFT functions  along with that of the equation of state parameter, $w_{\rm DE}$, which fixes the expansion history. 
Then, the underlying theory is fully specified~\cite{Hu:2014oga}.

We employ the  DE equation of state given by the Chevallier-Polarski-Linder (CPL) parametrization~\cite{Chevallier:2000qy,Linder:2002et}: 
\be
 w_{\rm DE}(a)=w_0 + w_a(1-a) \,,
\ee
where $w_0$ and $w_a$ are constants and indicate, respectively, the value and the time derivative of $w_{\rm DE}$ today. According to this choice, the density of the DE fluid evolves as
\be
\rho_{\rm DE}(a) = 3 m_0^2 H_0^2 \Omega^0_{\rm DE} a^{-3(1+w_0+w_a)}e^{-3 w_a (1-a)} ,\label{parametrization_rho_DE}\,
\ee
where $\Omega^0_{\rm DE}$ is the density parameter of  DE today.

\noindent
For the functional forms of the EFT functions we choose the following cases:
\begin{itemize}
\item \textbf{M1a}:
\ba
\Omega(a)=\Omega_0 a^{s_0}\,, \qquad \gamma_i(a)=0\,,\label{parametrization_M1}
\ea
where $\{s_0, \Omega_0\}$ are the constant parameters defining the $\Omega$ function.\\
\item \textbf{M1b}:
\ba
\Omega(a)=\Omega_0 a^{s_0}\,, \qquad \gamma_i(a)=\gamma_i^0a^{s_i}\,,\label{parametrization_M1}
\ea
where $\{s_i,\gamma_i^0\}$ are the parameters defining $\gamma_i$, with $i=1,2$. \\
\item \textbf{M2a}:
\ba
&&\Omega(a)=\Omega_0 a^{-3(1+w_0+w_a)}e^{-3 w_a (1-a)} \,, \nonumber\\
&& \gamma_i(a)=0 ,\label{parametrization_M2}
\ea
where $\Omega_0$ is a constant. This parametrization follows the DE density behaviour, as shown in eq.~\eqref{parametrization_rho_DE}.
\item 
\textbf{M2b}: 
\ba
&&\Omega(a)=\Omega_0 a^{-3(1+w_0+w_a)}e^{-3 w_a (1-a)} \,, \nonumber\\
&& \gamma_i(a)=\gamma_i^0 a^{-3(1+w_0+w_a)}e^{-3 w_a (1-a)} ,\label{parametrization_M2}
\ea
where $\gamma_i^0$  ($i=1,2$) are constants. \end{itemize}

We now focus on $\gamma_1$ and in particular on the its effects on the observables. As illustrated in the previous Section, in the QS limit $\gamma_1$ does not appear in either $\mu$ or  $\Sigma$, thus it is hard to know a priori which is the role it plays at the perturbations level. In ref.~\cite{Bellini:2015xja}, in the context of the $\alpha$-basis, it has been shown that the kinetic function $\alpha_K$, when parametrised as a function of the DE density parameter on a $\Lambda$CDM background, is hard to constrain with cosmological data. We thus expect a similar result for $\gamma_1$, since the two functions are related~\cite{Bellini:2014fua}.

For our study we consider the~\textbf{M1a} model and then we solely add $\gamma_1$, parametrised as in eq.~(\ref{parametrization_M1}). We compute the difference $\Delta C^{\rm TT}(\ell)$ between the Temperature-Temperature power spectra for the two models, in units of cosmic variance $\sigma_\ell = \sqrt{2/(2 \ell +1)} C^{\rm TT}(\ell)$, where for the latter $C^{\rm TT}(\ell)$ is the power spectra of the model with $\gamma_1=0$. We perform such procedure for a sample of $\sim 10^3$ models and we plot the results in Fig.~\ref{Cl_M1}. In such sample we have varied the background parameters in the ranges $w_0 \in [-1.5,0]$, $w_a \in [-1,0.5]$, the EFT functions parameters $\Omega_0 \in [0,3]$, $s_0 \in [0, 3]$, $\gamma_1^0 \in [0, 3]$ and $s_1 \in [-3,3]$. Let us note that these ranges have been chosen requiring the viability of the model against ghost and gradient instabilities~\cite{DeFelice:2011bh,Bellini:2014fua,Gergely:2014rna,Kase:2014yya,Kase:2014cwa,Gleyzes:2015pma,DeFelice:2016ucp,Frusciante:2017nfr,Frusciante:2018vht}. 

Analogously, in Fig.~\ref{Cl_M2} we plot the deviations in $C^{\rm TT}(\ell)$ when both $\Omega$ and $\gamma_1$ are parametrised as in {\bf M2}, eq.~\eqref{parametrization_M2}, considering the combinations $\{\Omega,\gamma_1=0\}$ and $\{\Omega,\gamma_1\}$. In this case we consider a similar sample of $\sim 10^3$ models, where $w_0$, $w_a$, $\Omega_0$ and $\gamma_1^0$ are varied in the same ranges as in  previous case.

From Figs.~\ref{Cl_M1} and~\ref{Cl_M2} we can infer that the effects of $\gamma_1$ on the TT power spectrum become significant for $\ell \lesssim 100$, due to the late-time ISW effect.  However, such contributions are always within the cosmic variance limit: we find that they never exceed $40\%$ and $90\%$ of comic variance  for {\bf M1} and {\bf M2} respectively. For this reason we conclude that it is unlikely that present surveys can constrain $\gamma_1$ or even that next generation experiments will gain constraining power on such operator.  We showed the results for the TT power spectrum, while we checked that for other observables we get similar results.
Nevertheless, $\gamma_1$ still plays an important role in the stability criteria of Horndeski theories. This means that, even if it does not directly modify the cosmological observables in a sizeable way, $\gamma_1$ has a strong effect on the allowed parameter space for the other EFT functions (see refs.~\cite{Bellini:2015xja,Kreisch:2017uet} for the analogous case of $\alpha_K$). In particular, it enters in the condition for the avoidance of ghost in the scalar sector~\cite{DeFelice:2016ucp}. 

\begin{figure}[t!]
\includegraphics[width=0.5\textwidth]{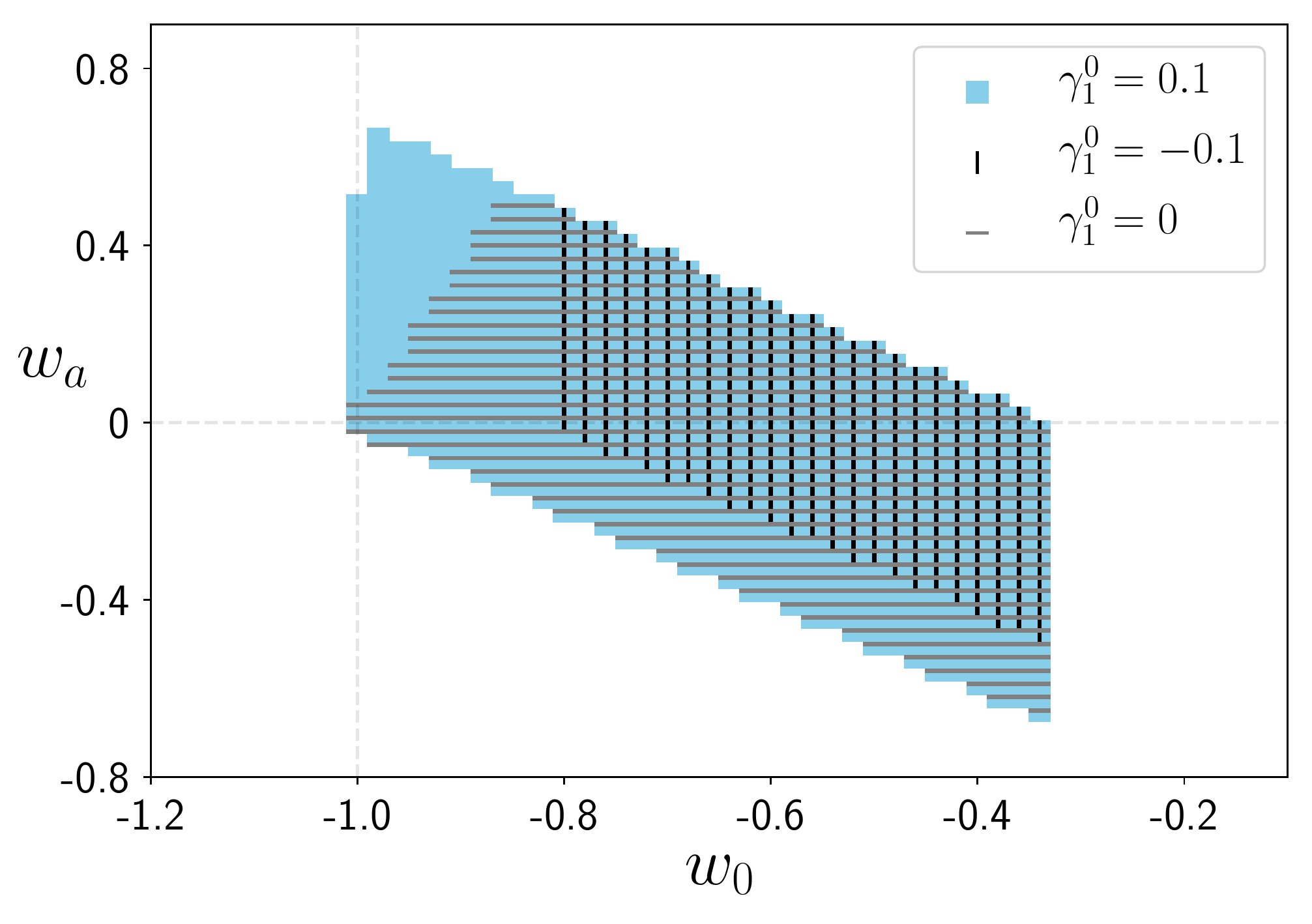}
\caption{\label{fig:CPL_stability} Effects of $\gamma_1$ on the stable CPL parameter space. We consider the parametrization of $\gamma_1$ defined in eq~\eqref{parametrization_M2} and compute the parameter space allowed by stability conditions, for different values of $\gamma_1^0$. The blue region represents the stable parameter space when $\gamma_1^0=0.1$, the horizontal grey lines refers to the case $\gamma_1^0=0$ and the vertical black lines to $\gamma_1^0=-0.1$.
}
\end{figure}

As an illustrative example of the relevance of $\gamma_1$ in the stability, we consider the model described solely by $\gamma_1$ ($\{\Omega,\gamma_2\}=0)$, when $\gamma_1$ is parametrised as in eq.~\eqref{parametrization_M2} on a CPL background.  We show in Fig.~\ref{fig:CPL_stability}  how drastically $\gamma_1$ changes  the  stable $w_0-w_a$ parameter space, for different values of $\gamma_1^0$. We see that changing the value of the latter parameter has a clear impact on the stability of the CPL parameters: a positive value enlarges the stable parameter space, while a negative $\gamma_1^0$ shrinks it.
Thus, we conclude that although $\gamma_1$ does not give any sizeable effect on the observables, it can not be neglected from the cosmological analysis, because of its important role in the stability conditions.
Moreover, as already pointed out in Sec.~\ref{Horndeskiparameterization}, when $\gamma_2\neq 0$ immediately follows  $\gamma_1\neq 0$. 
For this reason it is worth including such EFT function in the present cosmological analysis.  

\subsection{Codes and data sets}\label{Sec:Data}

For the present analysis we employ the EFTCAMB/EFTCosmoMC codes~\cite{Hu:2013twa,Raveri:2014cka,Hu:2014oga} \footnote{Web page: \url{http://www.eftcamb.org}}. The reliability of EFTCAMB has been tested against several Einstein-Boltzmann solvers and the agreement reaches the subpercent level~\cite{Bellini:2017avd}. 

We analyse \textit{Planck} measurements~\cite{Aghanim:2015xee,Ade:2015xua} of CMB temperature on large angular scales, i.e. $\ell < 29$ (low-$\ell$ likelihood), the CMB temperature on smaller angular scales, $30<\ell<2508$ (PLIK TT likelihood) and the CMB lensing map~\cite{Ade:2015zua}. 
We  also include Baryonic Acoustic Oscillations (BAO) measurements from  \textit{BOSS DR12} ({\it consensus} release)~\cite{Alam:2016hwk}, local measurement of $H_0$~\cite{Riess:2016jrr} and Supernovae data from the {\it Joint Light-curve Analysis} ``\textit{JLA}'' Supernovae (SN) sample~\cite{Betoule:2014frx}. 
Along with the former data set, we consider measurements from weak gravitational lensing from the Kilo Degree Survey (\textit{KiDS}) collaboration~\cite{Hildebrandt:2016iqg,Kuijken:2015vca,deJong:2012zb}. In this case we make a  cut at non-linear scales, by following the prescription in refs.~\cite{Kitching:2014dtq,Ade:2015rim}. Practically, one performs a cut in the radial direction $k\le 1.5 \, h$Mpc$^{-1}$ and one removes the contribution from the $\xi^-$ correlation function. In this way the analysis has been shown to be sensitive only  to the linear scales~\cite{Ade:2015rim}. 

We list the flat priors used for the models parameters presented in the previous Section:  $w_0\in[-5,0]$,  $w_a \in [-2,4]$ and $\{\Omega_0, s,\gamma_1^0,s_1,\gamma_2^0,s_2\} \in [-10,10]$.

\subsection{Forecast analysis}\label{Sec:Forecasts}

We use the Fisher matrix 
approach~\cite{tegmark_measuring_1998,seo_improved_2007,seo_baryonic_2005},
which is an inexpensive way of approximating the curvature of the likelihood at the peak, under the assumption that it is a Gaussian function of the model parameters. 
The main cosmological observables of next generation galaxy redshift surveys, such as \textit{Euclid}\footnote{http://www.euclid-ec.org/}~\cite{Amendola:2016saw, laureijs_euclid_2011},~\textit{DESI}\footnote{https://www.desi.lbl.gov/}~\cite{desi_collaboration_desi_2016-1,desi_collaboration_desi_2016}, \textit{LSST}\footnote{https://www.lsst.org/}\cite{2018arXiv180901669T} and \textit{SKA}\footnote{https://www.skatelescope.org/}~\cite{santos_hi_2015,Raccanelli:2015hsa,Bull:2015nra,Bull:2018lat}, are Galaxy Clustering (GC) and Weak Lensing (WL). WL can be measured with photometric redshifts and galaxy shape (ellipticity) data, while GC needs the position of galaxies in the sky and their redshifts to yield a 3-dimensional map of the large scale structure of the Universe. 
Though photometric GC can also give us some complementary information, especially in cross-correlation with WL, we use here only the more precise spectroscopic GC probe, which we assume to be independent of WL observables.
This is a rather conservative approach, meaning that our constraints might be weaker than in the full case with cross-correlations, as it has been shown with present surveys such as \textit{DES} \cite{PhysRevD.98.043526}.
Moreover,  we do not have a generally valid approach to calculate the non-linear matter power spectrum for models within the EFT formalism, thus we can not include non-linear scales in our modelling of the Fisher matrix. Therefore, we need to limit ourselves to linear scales, which might yield to large forecasted errors, especially for the WL analysis which is very sensitive to non-linearities. 
In practice, the largest scales we take into account correspond to  $k_{\rm min}=0.0079\textrm{h/Mpc}$ and, since we want to restrict ourselves to linear scales, we use a hard cut-off at $k_{\rm max} = 0.15 \textrm{h/Mpc}$ and at a maximum multipole of $\ell_{max}=1000$.
Finally, we  perform the forecast analysis only for the cases without massive neutrinos for the following reasons: firstly, we cut our analysis at non-linear scales and that is the regime where the larger effects coming from the presence of the neutrinos are expected; secondly, the results we get from cosmological data show that massive neutrinos do not affect considerably the constraints (see Sec. \ref{Sec:Results}).

\subsubsection{Galaxy Clustering}

In order to compute the predictions for Galaxy Clustering, we need to compute $P_{\rm obs}$, which is the 
Fourier transform of the two-point correlation function of galaxy number counts in redshift space. 
The observed galaxy power spectrum follows the matter power spectrum of the underlying dark matter distribution $P(k)$ up to a bias factor $b(z)$ and some effects related to the transformation from configuration space into redshift space. 
We assume the galaxy bias to be local and scale-independent, though  modified gravity theories might in general predict a scale dependence~\cite{Desjacques:2016bnm}. 
To write down the observed power spectrum, we neglect other relativistic and non-linear corrections, and we follow ref.~\cite{seo_improved_2007}, so that we end up with
\beq
\label{eq:observed-Pk}
\Pobs (k,\tilde{\mu},z)= \frac{D_{A,f}^{2}(z)H(z)}{D_{A}^{2}(z)H_{f}(z)} B^2(z) e^{-k^{2}\tilde{\mu}^{2}\sigma_{tot}^2}P(k,z)\,,
\eeq
with
\beq
\sigma_{tot}^2=\sigma_{r}^{2}+\sigma_{v}^{2}\,,\hspace{5mm}
B(z) = b(z) (1+\beta_{d}(z)\tilde{\mu}^{2})\, ,
\eeq
where $B(z)$ contains the so-called Kaiser effect \cite{eisenstein_cosmic_1999, kaiser_clustering_1987}, $\beta_{d}(z)\equiv f(z)/b(z)$, and 
$f\equiv d \ln G/d\ln a$ is the linear growth rate of matter perturbations.
In this equation  $\tilde{\mu}$, is the cosine of the angle between the line of sight and the 3d-wavevector
$\vec{k}$. Every quantity in this equation depends on all cosmological parameters and is varied accordingly, except for those with a subscript $f$, which denote an evaluation at the fiducial value.
$D_{A}(z)$ is the angular diameter distance 
and the exponential factor represents a damping term with
$\sigma^2_{r}+\sigma_v^2$, where $\sigma_r$ is the error induced by spectroscopic redshift 
measurements and $\sigma_{v}$ is the velocity dispersion associated to the Finger of God effect \cite{seo_improved_2007}. 
We marginalise over this last
parameter \cite{bull_extending_2015} and take a fiducial value $\sigma_{v} = 300$ km/s
compatible with the estimates in ref.~\cite{de_la_torre_modelling_2012}.  See refs.~\cite{CASAS201773,seo_improved_2007,Amendola:2011ie} for further details.

The Fisher matrix is then computed by taking derivatives of $P_{\rm obs}$ with respect to the cosmological parameters and by integrating these together with a Gaussian covariance matrix and a volume term, over all angles and all scales of interest \cite{CASAS201773, Casas-PhysRevD.97.043520}.

The galaxy number density $n(z)$ we use here, peaks at a redshift of  $z=0.75$ and it is similiar to the spectroscopic DESI-ELG survey found in \cite{desi_collaboration_desi_2016-1}.
We also use their expected redshift errors and bias specifications, but a slightly larger area of 15000 square degrees.

\subsubsection{Weak Lensing}

Weak Lensing is the measurement of cosmic shear, which represents the ellipticity distortions in the 
shapes of galaxy images. This in turn is related to deflection of light due to the presence of matter in the Universe. Therefore, WL is a very powerful probe of the distribution of large scale structures and due to its tomographic approach, provides valuable  information about the accelerated expansion of the Universe. Assuming small gravitational potentials and large separations, we can link cosmic shear to the matter power spectrum, giving direct constraints on the cosmological parameters.
In this case we use tomographic WL in which we measure the cosmic shear in a number of wide redshift bins, given by a window function $W_i(z)$ at the bin $i$, which is correlated with another redshift bin $j$. The width of these window functions depends on a combination of the photometric redshift errors and the galaxy number densities.
The cosmic shear power 
spectrum can thus be written as a matrix with indices $i, j$, namely
\beq
\label{def_shear}
C_{ij}(\ell)=\frac{9}{4}\int_{0}^{\infty}\textrm{d}z\frac{W_{i}(z)W_{j}(z)H^{3}(z)\Omega_{m}^{2}(z)}{(1+z)^{4}} \Sigma^2 (k,z) P_{m} \,,
\eeq
with $P_{m}$ evaluated at the scale $\ell/r(z)$, where the comoving distance is $r(z)$. 
In modified gravity, the lensing equation is modified by the term $\Sigma$ in eq.~(\ref{mudef}), thus  it turns out that such term also appears into the evaluation of the power spectrum.
For the Fisher matrix we follow the same procedure as in ref.~\cite{CASAS201773,Casas:2017eob}, 
where for the actual unconvoluted galaxy distribution function we have assumed
\begin{equation}
n(z)\propto (z/z_0)^{2}\exp\left(-(z/z_{0})^{3/2}\right) \, , \label{eq:ngal dist}
\end{equation}
and SKA2-like specifications for weak lensing \cite{harrison_ska_2016}, which despite 
being a rather futuristic survey, we decided to use here in order to improve our WL constraints, since we only deal with linear scales, which lowers a lot our constraining power.

\begin{table}[t!]
    \begin{tabular}{| l | c | c |  c | c | c | c | c |}
    \hline
    \Tstrut
    	Model 						&  $10^9 \, A_s$  &  	$n_s$ &		$\Omega_m$       		 & $H_0$           & $\Sigma m_\nu$  \\ \hline  \hline 
    	\Tstrut
	$\Lambda$CDM	     			& $  2.11^{+0.12}_{-0.12}   $& $  0.969^{+0.009}_{-0.009}   $ 	& $   0.297^{+0.013}_{-0.013}   $ 		& $   68.7^{+1.1}_{-1.0}    $ 	&     \\
	$\Lambda$CDM+$\nu$ 			& $  2.22^{+0.23}_{-0.19}    $ & $  0.974^{+0.012}_{-0.011}   $	& $   0.300^{+0.015}_{-0.014}   $ 		& $    68.4^{+1.2}_{-1.2}    $ 	& $  < 0.288  $  \\ \hline 
	\Tstrut
	M1a	             			& $  2.21^{+0.21}_{-0.21}    $&  $  0.974^{+0.012}_{-0.012} $ 	& $0.295^{+0.017}_{-0.016}   $ 		& $68.7^{+1.8}_{-1.7}        $ 	  &   \\
	M1a $+ \nu$    			&  $ 2.29^{+0.25}_{-0.22} $ & $  0.976^{+0.013}_{-0.013}   $	& $   0.298^{+0.017}_{-0.018}   $ 		& $68.4^{+1.8}_{-1.6}        $ 	& $< 0.281$  \\
	M1b	       			        & $  2.19^{+0.24}_{-0.23}  $ & $  0.973^{+0.013}_{-0.012}   $ & $0.293^{+0.017}_{-0.017}   $ 		&  $68.9^{+1.8}_{-1.8}        $ 	  &   \\
	M1b $+ \nu$    		        & $  2.28^{+0.25}_{-0.25} $ & $   0.975^{+0.013}_{-0.015}   $	& $0.295^{+0.018}_{-0.016}   $  		& $68.8^{+1.8}_{-1.7}        $ 	& $< 0.347  $   \\ \hline
	\Tstrut
	M2a 	             			& $ 2.27^{+0.21}_{-0.20}  $ 	& $  0.972^{+0.010}_{-0.010}   $& $0.302^{+0.015}_{-0.014}   $		& $68.1^{+1.3}_{-1.4}        $ 	  &   \\
	M2a $+ \nu$    			&  $  2.35^{+0.24}_{-0.22}  $ & $   0.975^{+0.011}_{-0.011}  $	&$0.303^{+0.016}_{-0.014}   $ 		& $67.9^{+1.3}_{-1.4}        $  	& $< 0.236 $    \\
	M2b	        				 &  $  2.20^{+0.28}_{-0.26}   $& $  0.968^{+0.013}_{-0.013}   $	& $0.300^{+0.016}_{-0.016}   $ 		& $68.6^{+1.8}_{-1.6}        $  	  &  \\
	M2b $+ \nu$     			 &  $  2.30^{+0.29}_{-0.29}  $ & $   0.970^{+0.014}_{-0.014}   $	& $0.304^{+0.017}_{-0.017}   $ 		& $68.5^{+1.7}_{-1.6}        $ 	& $< 0.543  $    \\
    \hline
     \end{tabular}
     \caption{\label{tab_bestfit_params1} 2$\sigma$ marginalised constraints on cosmological parameters. These values are obtained through the analysis of the full data set presented in Sec.~\ref{Sec:Data}.}  
\end{table}

\section{Results}\label{Sec:Results}

%
\begin{table*}[t!]
    \centering
    \begin{tabular}{| l | c | c | c | c | c | c |c |c | }
    \hline
    \Tstrut
    	Model 						&  $w_0$ 				& $w_a$ 				& $\Omega_0$ 		  & $s_0$& $\gamma_1^0$ & $s_1$& $\gamma_2^0$ & $s_2$ \\ \hline  \hline 
    	\Tstrut
	M1a  	             			&  $-1.04^{+0.14}_{-0.16}     $ 	& $0.22^{+0.46}_{-0.39}      $ 	&$-0.07^{+0.17}_{-0.18}     $ & $> 0.435 $ & & & &   \\
	M1a $ + \nu$    			&$-1.02^{+0.13}_{-0.18}     $	&  $0.12^{+0.49}_{-0.37}      $ 	& $-0.04^{+0.15}_{-0.21}     $ & $> 0.240$ & & & &   \\
	M1b 	         &  $-1.07^{+0.15}_{-0.16}     $	& $0.30^{+0.47}_{-0.42}      $ 	& $0.03^{+0.31}_{-0.25}      $ & $> 0.215                   $ & $> 0.217                   $&$  -- $& $-0.9^{+1.3}_{-2.0}        $& $ > 0.330  $\\
	M1b $+ \nu$      &  $-1.08^{+0.16}_{-0.15}     $ 	& $0.24^{+0.49}_{-0.48}      $ 	& $0.01^{+0.33}_{-0.33}      $ & $> 0.296 $ & $> 0.103                   $&$ -- $& $-1.9^{+2.3}_{-5.0}        $ & $> 0.147$\\ \hline
	\Tstrut
	M2a  	             			&  $-0.946^{+0.090}_{-0.060}  $ 	& $-0.098^{+0.25}_{-0.28}    $ 	& $0.018^{+0.032}_{-0.019}   $ &  			& & & &  \\
	M2a $ + \nu$    			&  $-0.950^{+0.087}_{-0.056}  $ 	&$-0.11^{+0.23}_{-0.30}     $  	& $0.019^{+0.037}_{-0.020}   $  &  & & & &     \\
	M2b &  $-0.94^{+0.15}_{-0.13}     $	& $-0.31^{+0.48}_{-0.63}     $ 	&  $0.047^{+0.068}_{-0.051}   $ &   & $  > 0.295  $&  &$-0.23^{+0.26}_{-0.32}     $ & \\
	M2b $+ \nu$      &  $-0.93^{+0.15}_{-0.14}     $ 	& $-0.61^{+0.66}_{-0.66}     $  	&  $0.080^{+0.099}_{-0.081}   $ &  & $  > 0.151 $& & $-0.36^{+0.36}_{-0.47}     $ &  \\
    \hline
     \end{tabular}
     \caption{\label{tab_bestfit_params2} 2$\sigma$ marginalised constraints on model parameters. These values are obtained through the analysis of the full data set presented in Sec.~\ref{Sec:Data}. $--$ means that the parameter is left unconstrained. }  
 \end{table*}
On top of the mentioned variety of  gravity models, we also consider two different cosmological scenarios: one with massless neutrinos and the other with a massive neutrino component.
In Tab.~\ref{tab_bestfit_params1} we show the results for the cosmological parameters for the models \textbf{M1a/b}, \textbf{M2a/b}, with and without massive neutrinos. In the same table we added, for comparison, the $\Lambda$CDM results. In Tab.~\ref{tab_bestfit_params2} we show the constraints on the corresponding model parameters. 

We also studied the effects of giving different hierarchies to the massive neutrinos species, considering the \textit{normal} (NH),  \textit{inverted} (IH) and \textit{degenerate} (DH) hierarchy scenarios. The impact of different hierarchies on cosmological constraints was first considered both in $\Lambda$CDM~\cite{Hannestad:2016fog,Simpson:2017qvj,Vagnozzi:2017ovm} and alternative cosmologies~\cite{2017PhRvD..95j3522Y,Peirone:2017vcq} and it is expected that the probability of breaking the degeneracy between them increases as the bound on the total mass of neutrinos becomes tighter~\cite{Hannestad:2016fog}. Nevertheless, we find that such different scenarios are indistinguishable when using this combination of data. The reason can be found in the following argument: in order to get any insight on a preferred hierarchy, one should get a sensitivity on the sum of neutrino masses  of  $\Sigma m_\nu < 0.2$~eV at 2$\sigma$, in particular to exclude the IH it has to be $\Sigma m_\nu< 0.1$~eV, as discussed in ref.~\cite{Hannestad:2016fog}. For the data sets and models considered in the present work, $\Sigma m_\nu$ never goes below this threshold at 2$\sigma$, (see Tab.~\ref{tab_bestfit_params1}).  

We find that, regardless of the model considered, the cosmological parameters $\{A_s, n_s, H_0, \Omega_m,\Sigma m_\nu\}$ are all consistent with the $\Lambda$CDM scenario at $2\sigma$. 
Furthermore, we do not find relevant differences when considering different combinations of the data sets, for such reason, we only show the results for the full data set analysis. 
Such constraints are not considerably affected by the presence of massive neutrinos or by the modifications to gravity introduced through $\Omega, \gamma_1$ and $\gamma_2$.  Finally, as shown in Tab.~\ref{tab_bestfit_params2}, $\gamma_1$ is really weakly constrained by the data and the results are mostly compatible with the prior we used. The cut in the negative prior range is due to the requirement of avoiding  ghost instability which enforces a positive $\gamma_1^0$. The same happens for the exponent parameter $s_1$ which is left totally unconstrained. This result is expected and in line with the discussion presented in Sec.~\ref{Sec:Model}.
\begin{figure}
	\includegraphics[width=0.5\textwidth]{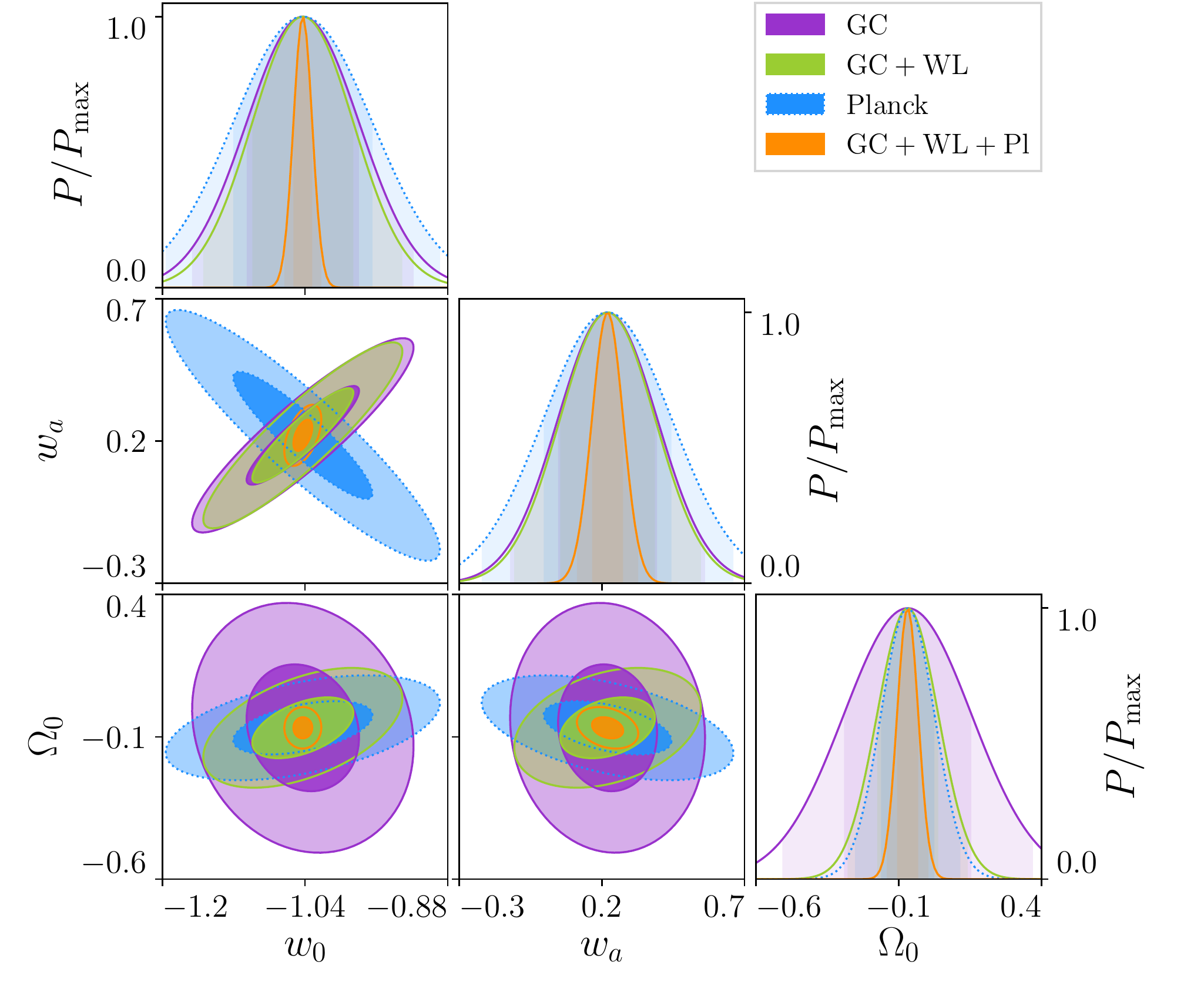}
	\caption{\label{M1OmFisher} Forecast for model {\bf M1a} for the equation of state parameters $w_0$, $w_a$ and the model parameter $\Omega_0$. In purple we have GC, in green GC+WL (assuming they are independent), in blue the Planck prior from our MCMC's and in orange the combination of GC+WL+Planck.
	}
\end{figure}
\begin{figure}
	\includegraphics[width=0.5\textwidth]{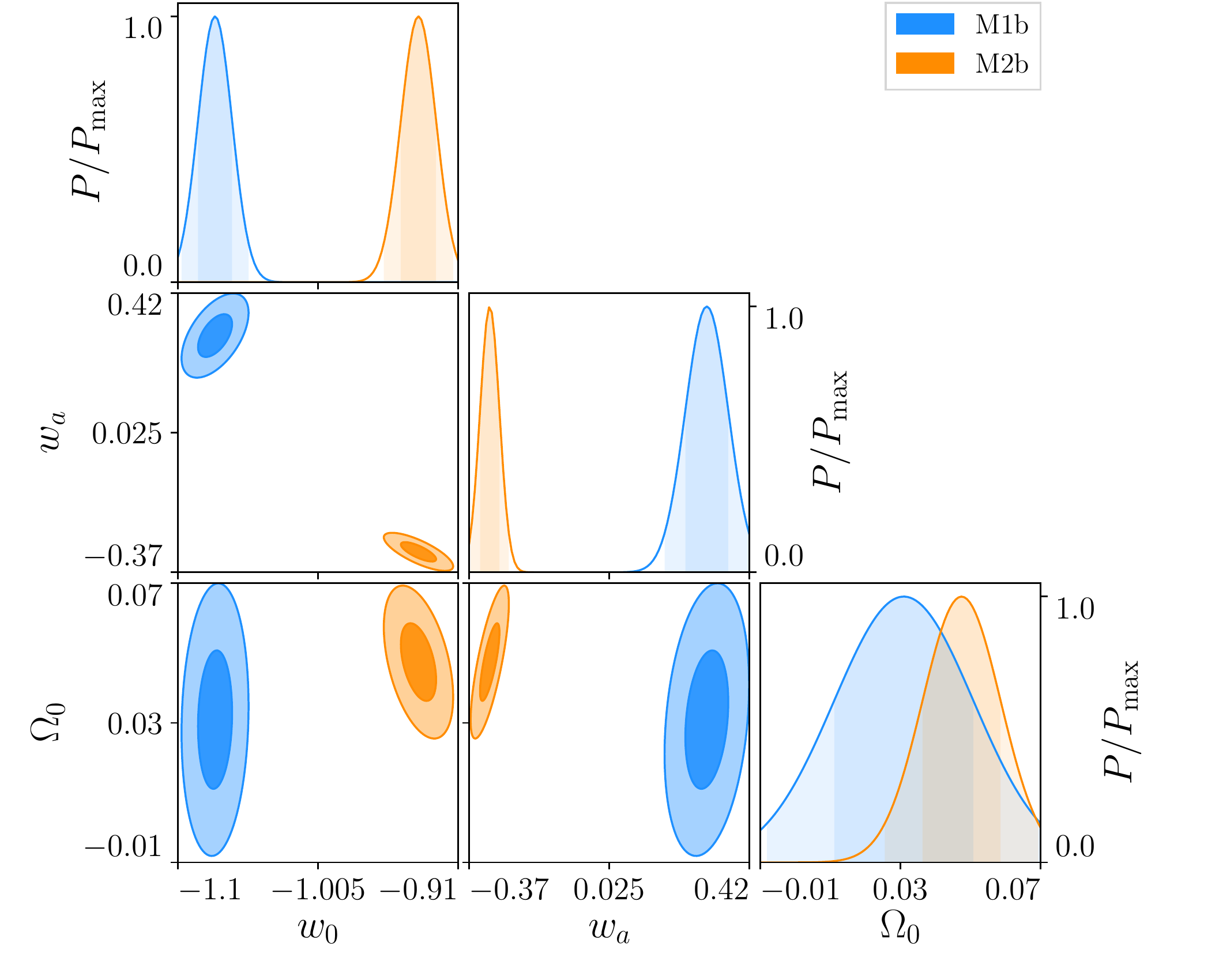}
	\caption{\label{M1bM2bFisher} Forecasts comparing model {\bf M1b} (blue) and {\bf M2b} (orange) for the equation of state parameters $w_0$, $w_a$ and the model parameter $\Omega_0$. Both Fisher matrices are computed for the combined GC+WL+Planck case.
	}
\end{figure}
\begin{figure}
	\includegraphics[width=0.5\textwidth]{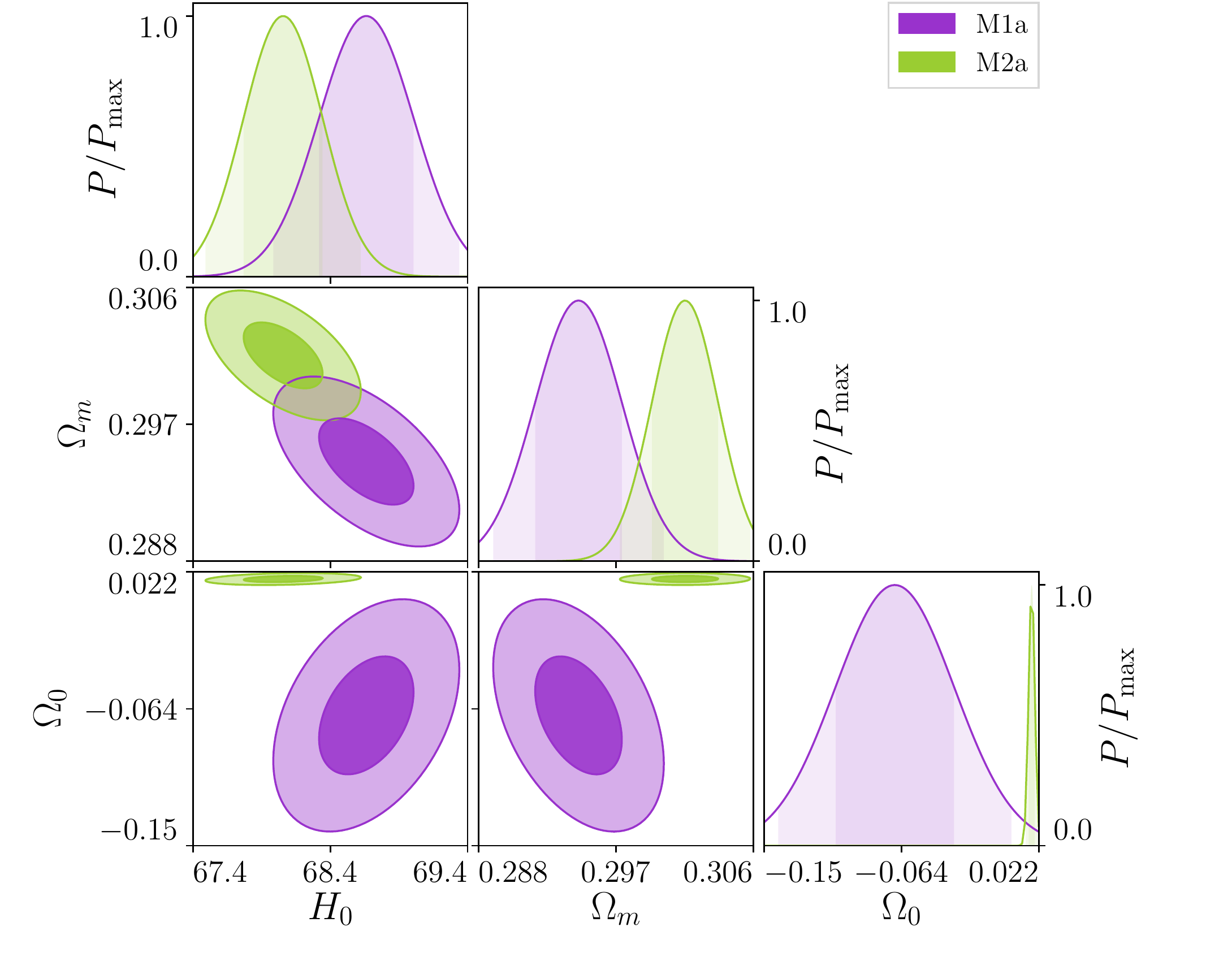}
	\caption{\label{M1aM2aFisher} Forecasts comparing model {\bf M1a} (purple) and {\bf M2a} (green) for the model parameter $\Omega_0$ and the cosmological parameters $H_0$ and $\Omega_m$. Both Fisher matrices  in this plot are computed for the combined GC+WL+Planck case.
	}
\end{figure}
\begin{table}[t!]
	\begin{tabular}{| l | c | c | c | c |}
		\hline
		\Tstrut
		Model & $2\sigma(10^9 A_s)$  & $2\sigma(\Omega_m)$ & $2\sigma(H_0)$ & $2\sigma(n_s)$    \\ \hline 
		 \hline 
		 \Tstrut
		M1a	             				&    4.0\% 	&   1.9\%	& 1.0\%   &  0.8\%     \\
		M1b	         	                &    4.2\%  	&   2.2\%  	& 1.1\%   &   0.9\%  \\
		M2a 	             		    &    0.02\%  	&   1.4\%		& 0.8\%   &   0.7\%  \\
		M2b	           	                &    4.4\%	&   1.7\%		& 1.0\%   &	0.8\%  \\
		\hline
	\end{tabular}
	\caption{\label{tab_forecast_std_pars} Forecasted $2\sigma$ errors on the cosmological parameters for a next generation spectroscopic Galaxy Clustering measurement plus a photometric Weak Lensing experiment, using Planck priors.}  
\end{table}
\begin{table}[t!]
	\centering
	\begin{tabular}{| l | c | c | c | c | c | c | }
		\hline
		\Tstrut
		Model 	&  $2\sigma(w_0)$ & $2\sigma(w_a)$ 	& $2\sigma(\Omega_0)$  & $2\sigma(s_0)$  &  $2\sigma(\gamma_2^0)$ & $2\sigma(s_2)$ \\ \hline  \hline 
		\Tstrut
		M1a  	&  $  2.0\% $ 	& $ 50\%  $ 	& $ 110\% $ &  $68\%$ & --    &  --	    \\
		M1b 	&  $  2.2\% $ 	& $ 40\%  $ 	& $ 128\% $ &  $96\%$ & $240\%$ &  $136\%$	   \\
		M2a 	&  $  1.9\% $ 	& $ 44\%  $ 	& $ 22\% $  &   --  &  --   &  --\\
		M2b     &  $  2.6 \%$	    & $ 18\%  $   & $ 48\% $ & -- & $40\%$      & -- \\
		\hline
	\end{tabular}
	\caption{\label{tab_forecast_modelparams} Forecasted 2$\sigma$ marginalized constraints on model parameters. These values are obtained with the combination of GC+WL+Planck for a future next generation galaxy survey. }  
\end{table}

Let us now move to the forecasts. For the fiducial parameters we use the best fits values from Tabs.~\ref{tab_forecast_std_pars} and ~\ref{tab_forecast_modelparams}. For the models with $\gamma_1$ we have used $\gamma_1^0 = 5.0$ and $s_1 = 1.4$ for \textbf{M1b},  while for \textbf{M2b}  we used $\gamma_1^0 = 4.4$.  We considered these values as fixed, since we proved that the effect of $\gamma_1$  is negligible on the cosmological observables, even for next generation surveys.

In Fig.~\ref{M1OmFisher} we show the forecasted $1$ and $2\sigma$ constraints, for the model parameters of \textbf{M1a}, for different combinations of the next generation datasets. 
From such plots we can see the effect of the different datasets: we find a common feature in the $w_0-w_a$ plane, where  the GC analysis removes the degeneracy coming from the Planck measurements; analogously, we can appreciate how the inclusion of WL in the CG analysis considerably increases the constraints on $\Omega_0$.

In Figs.~\ref{M1bM2bFisher}-\ref{M1aM2aFisher} we compare the forecasted marginalized distribution for the models \textbf{M1b}-\textbf{M2b} and \textbf{M1a}-\textbf{M2a} respectively, obtained through the analysis with the full CG+WL+Planck dataset. From these results we can see that the \textbf{M1b}-\textbf{M2b}  models have the fiducial values of $\Omega_0$ compatible within the error bars, while in the $w_0-w_a$ parameter space the models could be distinguished at more than $5\sigma$. 
Alternatively, in the \textbf{M1a}-\textbf{M2a} comparison plot, while the constraints on  cosmological parameters, $H_0$ and $\Omega_m$, are very similar, the constraint on $\Omega_0$ for the model \textbf{M2a} is much stronger (GC and Planck). This is due to the fact that in \textbf{M2a} the parameter $\Omega_0$ is related to $w_0$-$w_a$ and therefore can be measured indirectly by measuring the equation of state of dark energy. In the marginal likelihood of $\Omega_0$ both models could be distinguished at almost the 3$\sigma$ level.

In Tabs.~\ref{tab_forecast_std_pars}-\ref{tab_forecast_modelparams} we list the forecasted $2\sigma$ errors respectively on the cosmological and model parameters obtained with the GC+WL+Planck combination, for a future next generation galaxy survey. 
Compared to present data we find  that future surveys in general will slightly improve the constraint on cosmological parameters, notably  for the $A_s$ parameter in \textbf{M2a}  the error reduces by 2 orders of magnitude in the forecasts. Such improvement is due to the WL which breaks the degeneracy with CG and Planck.  Furthermore, future surveys  will improve the constraints on the models parameters by one order of magnitude. Even better they will set constraints of order $\lesssim 100\%$ on $s_0$, $s_2$ parameters for which the present data adopted in this work are able only to set lower bounds. 

We also explore the  deviations from GR of the $\mu$ and $\Sigma$ functions and we test the goodness of their QS approximations.
For these purposes we compare the QS expressions for $\mu$ and $\Sigma$, as reported in eqs.~(\ref{QS_expressions}), with those obtained by using their exact expressions as in eq.~(\ref{mudef}) (hereafter we will use the superscript ``ex''). These are computed by evolving the full dynamics of perturbations with EFTCAMB. Finally, we show the deviations of the exact solutions with respect to GR. The cosmological/models parameters are chosen accordingly to the bets fit values in Tabs.~\ref{tab_bestfit_params1}-\ref{tab_bestfit_params2}. We did not include the case of massive neutrinos, since their presence does not make any consistent difference. 

For the \textbf{M1a/b} models we find that the QS approximation is a valid assumption at the values of $z$ and $k$ considered, being the difference between $\mu/\Sigma$ QS and exact $\sim 10^{-3} (0.1\%)$,  and $\mu/\Sigma$ are also compatible with GR ($|\mu-1|$ and $|\Sigma-1| \sim 10^{-3}$). 

\begin{figure}[t!]
\centering
\includegraphics[width=0.5\textwidth]{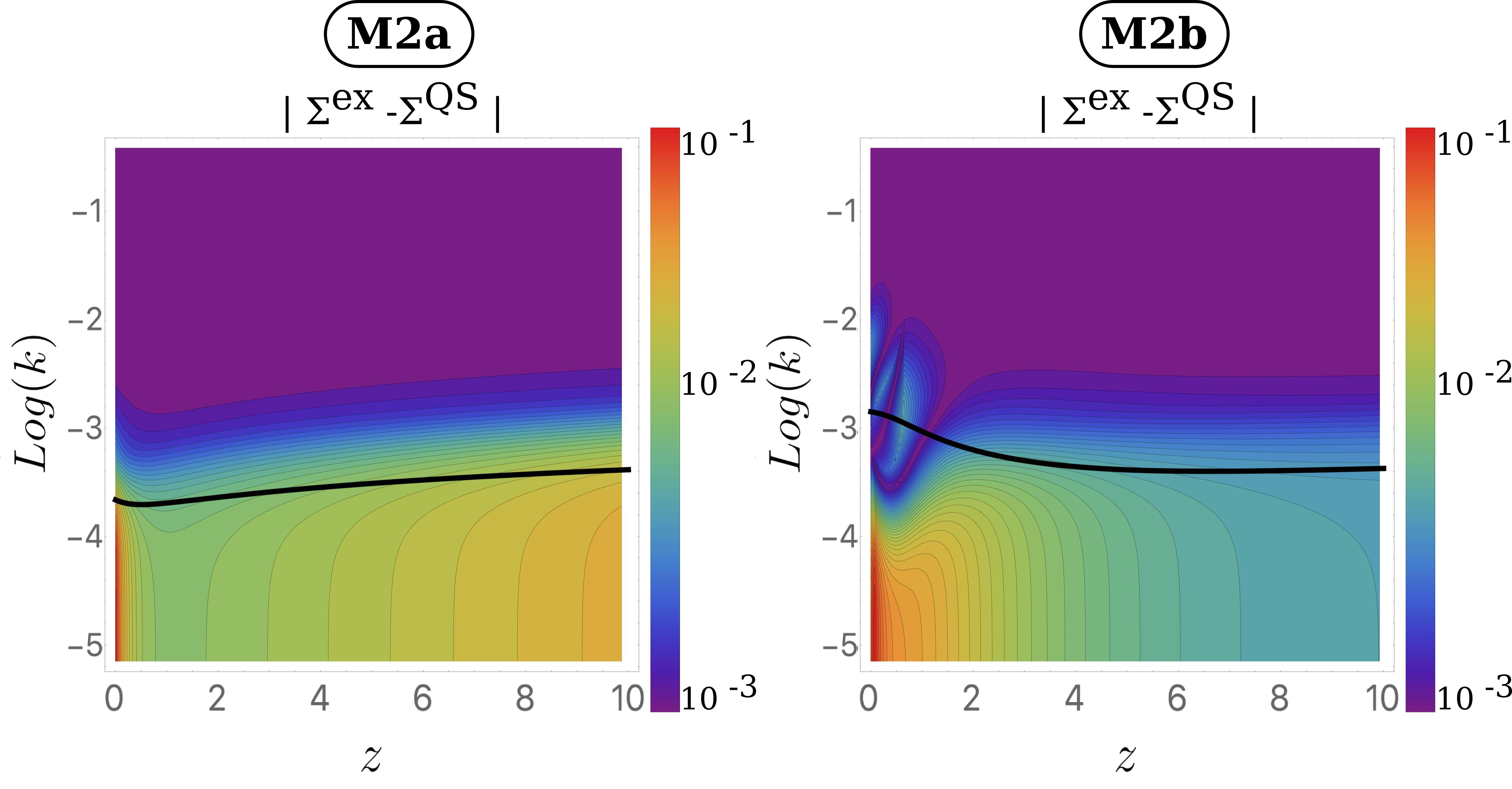}
\includegraphics[width=0.5\textwidth]{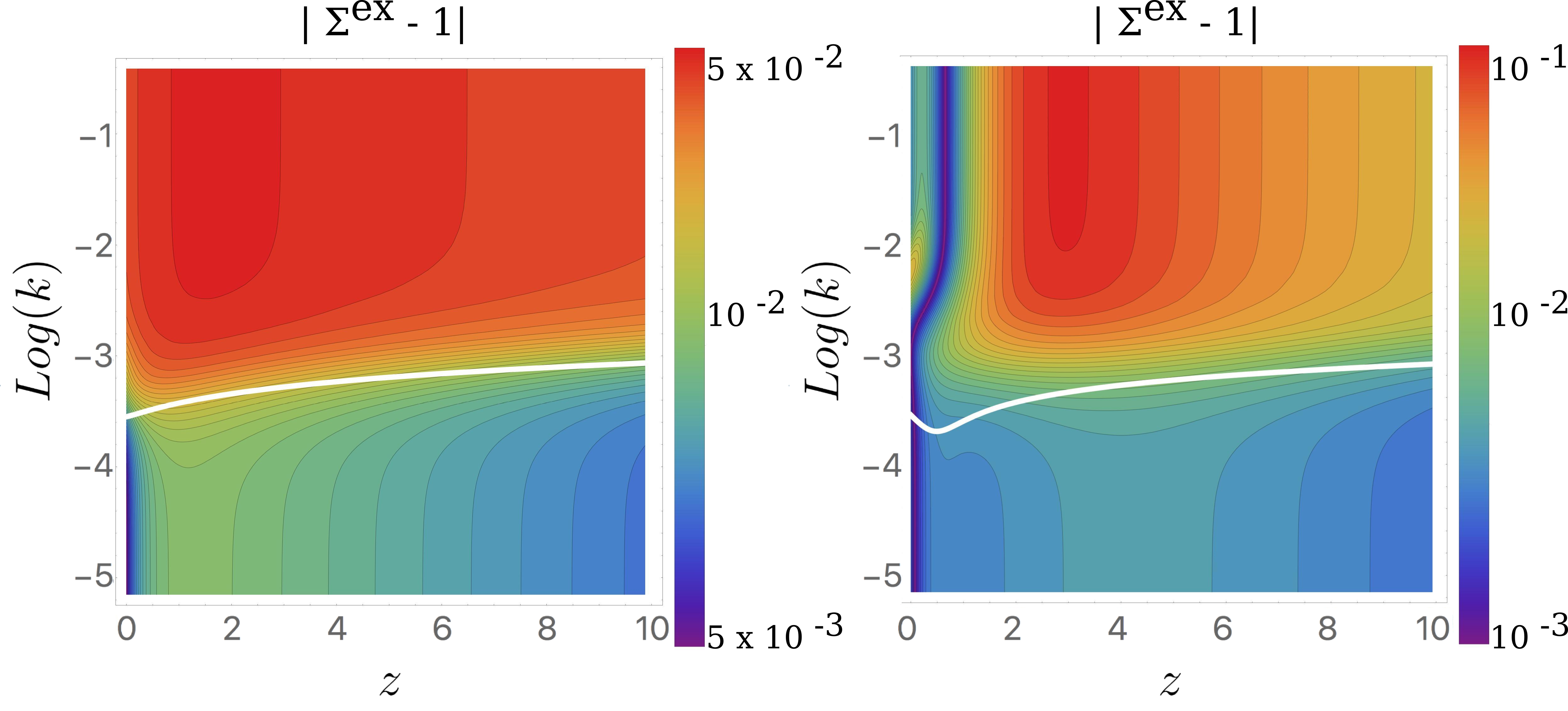}
\caption{\label{fig:M2_Sigma}  \textit{Top panels:} Absolute relative differences between the exact (``ex'') and QS calculations of $\Sigma$ for \textbf{M2a} model (top left) and  \textbf{M2b} (top right).
The black line corresponds to the dark energy sound horizon scale.   \textit{Bottom panels:}  Absolute relative differences between exact $\Sigma$ in models \textbf{M2a} (bottom left) and \textbf{M2b} (bottom right) and the General Relativity value. The white line corresponds to the Compton scale associated to the extra scalar DoF.  
In all panels we have used the best fit values for the parameters in Tabs.~\ref{tab_bestfit_params1} and~\ref{tab_bestfit_params2}. 
For reference, we checked that such results show a behaviour similar to what we obtain for the $\mu$ function.}
 \end{figure}

For the \textbf{M2a/b} cases we find different results, as we show in Fig.~\ref{fig:M2_Sigma}. 
In the top panels we plot the difference between the QS and exact solutions. We can see
that the QS approximation is a valid assumption within the sound horizon ($k_s=c_sk_H$, black line). Indeed, for both \textbf{M2a} and \textbf{M2b} the quantity $\Delta \Sigma=|\Sigma^{\rm ex}-\Sigma^{\rm QS}|$ reaches around $0.1\%$ deep inside the $k_s$, while outside it grows to few percents, reaching around $10\%$ at small $z$.  
Finally, we explore the deviations of \textbf{M2} model from GR. From the bottom panels in Fig.~\ref{fig:M2_Sigma}, one can clearly see that the Compton wavelength ($k_C$, white line) associated to the extra scalar DoF actually introduces a transition between two regimes. In fact, the large deviations from GR at $k>k_C$ can reach $5\%$ at all redshift (\textbf{M2a}) or for $z>1$ (\textbf{M2b}). On the other hand, at larger scales ($k<k_C$) $\Sigma$ gets closer to its GR value, with a relative difference which is always below the $1\%$. 
 
Such results are particularly interesting when we want to extend the forecasts and analyze the constraining power of future surveys on the phenomenological functions $\Sigma, \mu$.
Using the QS expressions in eqs.~(\ref{QS_expressions}) and the Fisher matrices obtained for the model parameters, we can calculate a derived Fisher matrix $\mathbf{\tilde F}$ for the forecasted errors on the derived quantities, $\mu$ and $\Sigma$ as follows 
\ba
\mathbf{\tilde{F}} =  \mathbf{J}^{T} \mathbf{F} \mathbf{J}  \,,
\ea
with
\ba
\mathbf{J} \equiv J_{ij} =  \frac{\partial p_i }{\partial \tilde{q}_j}
\ea
where $p_i$ is a vector containing all the parameters of the model (standard cosmological parameters ($\Omega_m$, $H_0$, $A_s$, $n_s$, $w_0$ and $w_a$) together with EFT parameters ($\Omega_0$, $\gamma_2^0$, $s_i$, ...)) and $\tilde{q}$ is a vector containing the standard cosmological parameters plus $\mu$ and $\Sigma$. 
Through the QS limit we can compute $ {\partial \tilde{q}_j}/{\partial p_i }$, since we know the functions $\Sigma(k,z,p_i)$ and $\mu(k,z,p_i)$. So, in order to compute the Jacobian $\mathbf{J}$ it can be shown that its inverse is equal to $J^{-1}_{ij} = {\partial \tilde{q}_j}/{\partial p_i }$.

\begin{figure}
	\centering
	\includegraphics[width=0.99\linewidth]{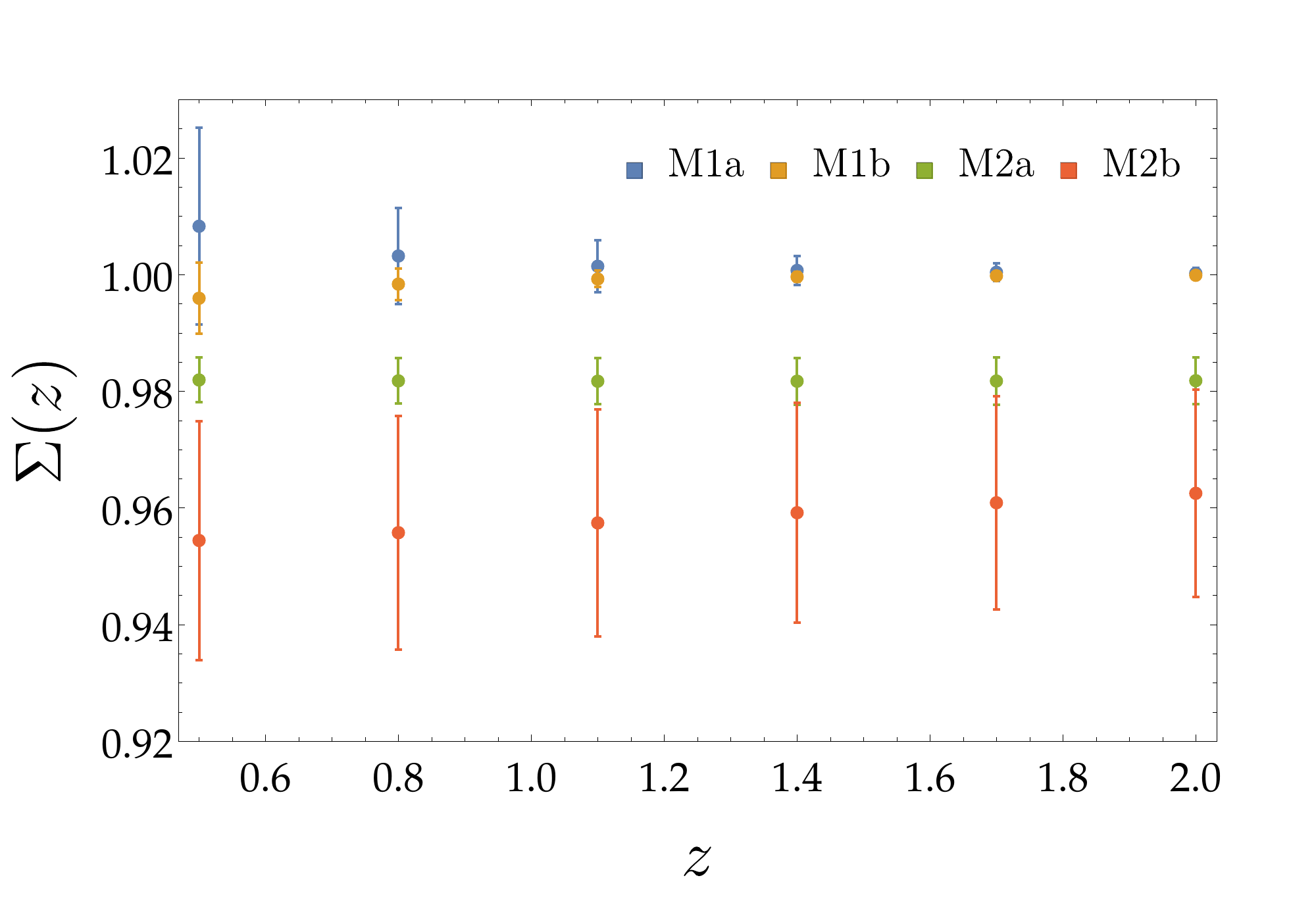}
	\caption{	\label{fig:mu-errors-allmodels}Forecasted 2$\sigma$ errors on the $\Sigma(z)$ parameter, for all four models considered in this work. The errors turn out to be quite small, since the parameter dependence of $\Sigma$ is very mild in the QS limit. We fix $k=0.01\,\, h/$Mpc.}
\end{figure}

We compute the derived Fisher matrices at a fixed scale, i.e. $k=0.01\,\, h/$Mpc which is well inside the Compton scale, for which the QS approximation is valid and where linear structure formation still holds. We do the same for 6 redshift bins, between $z=0.5$ and $z=2.0$, which cover typical redshift ranges of future surveys.
We report in Fig.~\ref{fig:mu-errors-allmodels} the $2\sigma$ error on $\Sigma(z)$ after marginalising over all other parameters.
We obtain the same errors for $\mu(z)$, since for our models this function behaves extremely similar to $\Sigma(z)$.


For models \textbf{M1a} and \textbf{M1b}, the errors obtained are of the order of $10^{-3}$ , decreasing towards $10^{-4}$ for higher redshifts ($z>1.5$), since there the functions $\mu(z)$ and $\Sigma(z)$ asymptotically tend to 1, independent of the cosmological parameters, which then implies very small predicted errors.
For models \textbf{M2a} and \textbf{M2b} the forecasted errors are constant in redshift, being approximately $4\times 10^{-3}$ and $2\times 10^{-1}$, respectively.

\section{Conclusion}\label{Sec:conclusion}

In this work we have explored the phenomenology of the class of Horndeski theory compatible at all redshift with the gravitational waves constraints, which we called surviving Horndeski (sH). For this class of modified gravity models we have  provided cosmological constraints from present day and upcoming large scale surveys. 
We performed the study by means of the EFT framework: thus we moved the problem of choosing the sH functions $\{\mathcal{K},G_3,G_4\}$ to selecting the free functions in the EFT formalism $\{\Omega,\gamma_1,\gamma_2\}$. For this particular class of models the mapping procedure becomes quite straightforward and there exists a one to one correspondence between each EFT function and the Horndeski ones. 

We found that the main contribution of the EFT function $\gamma_1$ dwells in the late-time ISW effect, but always within the cosmic variance limits. We could then infer that both present and future surveys can not constraint the evolution of $\gamma_1$: this is confirmed by the results of our cosmological analysis in Tab.~\ref{tab_bestfit_params2}, which left the $\gamma_1$ parameters completely unconstrained. However, let us note that the use of sophisticated multi-tracer techniques could allow to overcome the cosmic variance limitations~\cite{Camera:2016cpr}.
Moreover, we showed that $\gamma_1$ still has an important role in defining the viable parameter space of the theory, thus it can not be neglected in the cosmological analysis. 

We provided a constraint analysis of the sH models, using present day data and forecasts from  combinations of  GC and WL for a generic next generation galaxy survey. We found that future surveys will be able to increase the precision on the model parameters constraints by one order of magnitude.
In the forecast analysis  we did not notice any peculiarity at the level of the cosmological parameters, whose error bars are compatible among all models, we highlighted many features related to the model parameters for the single cases. For example, we were able to show that the correlation between $\gamma_2^0$ and $w_a$ in \textbf{M2b} translates in tighter constraints on the latter parameter. Furthermore, in Figs.~\ref{M1bM2bFisher} and~\ref{M1aM2aFisher} we showed the \textbf{M1b}/\textbf{M2b} and \textbf{M1a}/\textbf{M2a} model comparisons for the forecasted marginalized results. From such comparisons we are able to state that, given these fiducials, we will be able to distinguish \textbf{M1b} from \textbf{M2b} at $5\sigma$ level in the $w_0-w_a$ parameter space and, analogously, \textbf{M1a} from \textbf{M2a} at $3\sigma$ in the marginal likelihood of $\Omega_0$.

We studied the deviations of \textbf{M1} and \textbf{M2}, with respect to GR, in terms of the phenomenological functions $\mu$ and $\Sigma$. We found that \textbf{M1} is compatible with GR within $0.1\%$, while \textbf{M2a/b} show a $5\%$ departure from GR, at scales smaller than the Compton scale. We then tested the validity of the QS approximation and we found that it is a valid assumption for the \textbf{M1} model regardless of the scale, while, in the case of \textbf{M2}, we numerically checked that the validity of the QS limit is deeply connected with the definition of the dark energy sound horizon scale: within this scale the approximation holds at sub percent level, while it breaks down at larger scales. This result is in complete agreement with what found in ref.~\cite{Sawicki:2015zya}. 
Finally, we propagated the forecasted errors on the model parameters into $\mu$ and $\Sigma$ and we found that for models \textbf{M1a/b} the forecasted 2$\sigma$ errors, despite being very small ($\sim10^{-3}$) will not be able to discriminate these models from GR at more than 1$\sigma$, because both $\mu, \Sigma$  are  close to the  GR values, i.e.  $\mu=\Sigma=1$.
On the contrary, for models \textbf{M2a/b} the discrepancy to GR is large  and the errors are small enough, such that, provided the same best-fit values hold, we could be able to distinguish these models from standard GR at more than 3$\sigma$ in the derived quantities $\mu$ and $\Sigma$, using future galaxy surveys combined with CMB priors.

We conclude that the surviving class of Horndeski theory offers an interesting cosmological phenomenology, even after the $c_t^2=1$ constraint, and it is worth to be further investigated with the upcoming observational data.
Future surveys will provide a large amount of high precision-data, not only limited to the galaxy clustering and weak lensing observables considered here, and the inclusion in the data analysis of a proper treatment of non-linear scales will further improve their power in constraining~\cite{Casas:2017eob}.    
Such high sensitivity will set tiny constraints on any signature of deviations from GR allowing to discriminate among gravity models and it will represent the ultimate test for the $\Lambda$CDM scenario.

\acknowledgments

We thank Martin Kilbinger, Martin Kunz, Matteo Martinelli, Shinji Mukohyama, Valeria Pettorino and Alessandra Silvestri  for useful discussions and comments on this work.
The research of NF is supported by Funda\c{c}\~{a}o para a  Ci\^{e}ncia e a Tecnologia (FCT) through national funds  (UID/FIS/04434/2013), by FEDER through COMPETE2020  (POCI-01-0145-FEDER-007672) and by FCT project ``DarkRipple -- Spacetime ripples in the dark gravitational Universe" with ref.~number PTDC/FIS-OUT/29048/2017.
SP acknowledge support from the NWO and the Dutch Ministry of Education, Culture and Science (OCW), and also from the D-ITP consortium, a program of the NWO that is funded by the OCW.
SC acknowledges support from CNRS and CNES grants.
NF, SC and  SP  acknowledge the COST Action  (CANTATA/CA15117), supported by COST (European Cooperation in  Science and Technology).
NAL acknowledges support from DFG through the project TRR33 ``The Dark Universe'', and would like to thank the Department of Physics of the University of Lisbon for its hospitality during a week stay.

\bibliography{EFT_RPH}

\end{document}